\documentclass{ws-procs975x65}
\usepackage{graphicx}

\def\be{\begin{equation}}
\def\ee{\end{equation}}
\newcommand{\bea}{\begin{eqnarray}}
\newcommand{\eea}{\end{eqnarray}}
\newcommand{\nn}{\nonumber}

\begin{document}

\title{A public framework for Feynman calculations\\ and post-Newtonian gravity}

\author{Michele Levi}

\address{Institut de Physique Th\'eorique, CEA \& CNRS, Universit\'e Paris-Saclay\\
91191 Gif-sur-Yvette, France\\
michele.levi@ipht.fr}

\begin{abstract}
We report here on a line of work that has played a key role in formally establishing and going beyond 
the state of the art in the effective field theory (EFT) approach and in post-Newtonian (PN) gravity. 
We also outline here how this comprehensive framework in fact forms the outset of a prospective rich 
research program, building on the public Feynman and PN technology developed. 
\end{abstract} 

\keywords{Effective field theories; Post-Newtonian theory; Gravitational waves; High precision 
calculations.}

\bodymatter

\section{Introduction}

In recent years significant progress was made in a line of research, based on an Effective Field Theory 
(EFT) approach for the treatment of gravitational waves (GWs), emitted from coalescing binaries, which was put forward by Goldberger et al.\cite{Goldberger:2004jt,Goldberger:2007hy, 
Goldberger:2009qd}. Such binaries contain compact components, which still orbit in non-relativistic 
(NR) velocities in their long inspiral phase, and are therefore analytically studied via the 
post-Newtonian (PN) approximation in gravity\cite{Blanchet:2013haa}. Here we report in particular on a 
series of works, which focused on the treatment of the challenging and realistic case, where the 
constituents of the binary are rotating\cite{Levi:2008nh,Levi:2010zu,Levi:2011eq,Levi:2014sba, 
Levi:2015msa,Levi:2014gsa,Levi:2015uxa,Levi:2015ixa,Levi:2016ofk}, and on making the general EFT 
framework widely accessible and available to the community, including a public automated computation 
tool \cite{Levi:2017kzq,Levi:2017oqx,Levi:2018nxp}. This line of work has played a key role in 
formally establishing and going beyond the state of the art in the EFT approach and in PN gravity, 
where we also outline here how this comprehensive framework in fact forms the outset of a prospective 
rich research program, building on the public Feynman and PN technology developed. For a general 
overview presentation of the main unique advances in the field of EFTs of PN gravity, and the various 
prospects of using field theory to study gravity theories at all scales, see Ref.~\refcite{Levi:2019rli}.

\section{Effective Field Theories Of Post-Newtonian Gravity}

Multi-scale problems are prevalent in physics, and as it turns out the compact binary inspiral is also 
of this kind. Such problems are amenable to a systematic treatment using the framework of EFTs, 
originally developed in the context of quantum field theories (QFTs) and particle physics. The compact 
binary inspiral involves indeed a hierarchy of scales, controlled by the NR orbital velocity, $v$, 
which constitutes the small parameter of the effective theories at each of the intermediate scales in 
the problem. There are three widely separated characteristic scales in the binary inspiral problem: 
The scale of the single compact object, $r_s$, the scale of the orbital separation of the binary, $r$, 
and the scale of radiation, emitted from the binary in GWs, with a wavelength $\lambda$. These scales 
satisfy $r_s\sim Gm \sim r v^2 \sim \lambda v^3$, where $m$ is the typical mass of the single compact 
object, $v\ll1$, and in the PN approximation $n$PN $\equiv v^{2n}$ order corrections to Newtonian 
gravity are found. Therefore one needs to proceed to construct a tower of EFTs in stages, where in 
each stage an effective theory is devised to remove the corresponding intermediate characteristic 
scale.

The setup of EFTs is then universal, and it follows one of the two complementary procedures to 
construct an EFT, which are referred to as ``bottom-up'' and ``top-down''. Both of these approaches are also used in the binary inspiral problem. At the first stage, where the scale of the single object is to be 
eliminated, the full theory of gravity is considered as General Relativity (GR), and is given, e.g.,~in 
terms of the following Einstein-Hilbert action: 
\be
S[g_{\mu\nu}]= -\frac{1}{16\pi G} \int d^4x \sqrt{g}\,R[g_{\mu\nu}], 
\ee
with $g_{\mu\nu}$ for the gravitational field. Then, in the bottom-up approach we construct the 
effective action by identifying the relevant degrees of freedom (DOFs) and the symmetries at the 
desirable scale. In this case, this means that the remaining bulk action, after decomposing the 
gravitational field into $g_{\mu\nu}\equiv g^s_{\mu\nu}+\bar{g}_{\mu\nu}$, where $g^s_{\mu\nu}$ 
represents the strong field modes to be removed in the effective action, is augmented with a worldline 
action, which contains new worldline DOFs of a point particle. The effective action is then of the 
following form:
\bea \label{oneparteft}
S_{\text{eff}}[\bar{g}_{\mu\nu}(x),y^\mu(\sigma),e^\mu_A(\sigma)]
&=&S[\bar{g}_{\mu\nu}] \,+\, 
S_{\text{pp}}[\bar{g}_{\mu\nu}(y),y^\mu(\sigma),e^\mu_A(\sigma)]\nn\\
&=&-\frac{1}{16\pi G}\int d^4x\sqrt{\bar{g}}\,R[\,\bar{g}_{\mu\nu}]
+\sum_{i=1}^{\infty} C_i(r_s) \int d\sigma \,{\cal{O}}_i(\sigma),
\eea
where the worldline DOFs depend on the worldline parameter $\sigma$, ${\cal{O}}_i$ are the operators evaluated on the worldline, and $C_i$ denote the ``Wilson 
coefficients'', which encapsulate the UV physics suppressed, and depend only on the scale $r_s$.

At the second stage, the field component is further decomposed into components with definite scaling 
in the velocity, i.e.~$\bar{g}_{\mu\nu}\equiv\eta_{\mu\nu}+H_{\mu\nu}+\widetilde{h}_{\mu\nu}$, where 
$H_{\mu\nu}$ denotes the orbital field modes, which we want to eliminate at this stage, and 
$\widetilde{h}_{\mu\nu}$ denotes the radiation modes. We then proceed with the top-down approach to 
integrate out the orbital scale by starting from the following two-particle action, obtained from the 
previous stage:
\be \label{twoparteft}
S_{\text{eff}}
[\bar{g}_{\mu\nu},y_1^\mu,y_2^\mu,e_{1\,A}^{\,\,\mu},e_{2\,A}^{\,\,\mu}]=
S[\bar{g}_{\mu\nu}] \,+\, 
\sum_{a=1}^{2}
S_{pp}[\bar{g}_{\mu\nu}(y_a),(y_a)^\mu,(e_a)_{A}^{\,\mu}](\sigma_a).
\ee
We integrate out the orbital field modes explicitly, using a standard diagrammatic expansion with 
the following definition for the new effective action:
\be \label{seff2part}
e^{iS_{\text{eff}}[\widetilde{h}_{\mu\nu},(y_c)^{\mu},(e_c)_{A}^{\,\mu}]}
\equiv
\int {\cal{D}} H_{\mu\nu}\,e^{iS_{\text{eff}}
	[\bar{g}_{\mu\nu}, y_1^\mu,y_2^\mu,e_{1\,A}^{\,\,\mu},e_{2\,A}^{\,\,\mu}]},
\ee 
where now there are new worldline DOFs, with the subscript `c', of a composite particle, that is the binary system. 

Finally, when the conservative sector is concerned, no radiation modes are present, and one should 
arrive at an effective action, which no longer contains field modes, or another scale to remove.

\section{Spinning Particles In Gravity}

Gravitating spinning objects have been considered as particles in an action approach both in special and general relativity already more than 40 years ago, see 
e.g.~Refs.~\refcite{Hanson:1974qy,Bailey:1975fe}, yet they were first tackled in the context of the EFT approach for PN computations of the binary inspiral in Ref.~\refcite{Porto:2005ac}. Spin in Relativity
necessarily implies having a non-vanishing finite size\footnote{Otherwise, the rotational velocity of 
the object would surpass the speed of light.}, i.e.~a deviation from the point particle view, which is 
also adopted in the EFT approach, and this poses a challenge for the construction of an EFT for a 
gravitating spinning particle. Moreover, once we leave Newtonian physics, a unique notion of a 
``center of mass'' is lost, and hence there is an ambiguity as to the choice of the variables 
describing the rotation of the object.

Nevertheless, as we noted above the way to proceed is to construct an EFT for a gravitating spinning 
particle, following a bottom-up procedure, i.e.~by identifying the relevant DOFS, and coupling them in 
all possible ways allowed by the proper symmetries at this scale. After this first EFT is obtained, we 
can proceed to explicitly compute, using the top-down procedure starting from the two-particle EFT, 
the PN binding interactions of the binary system, i.e.~the second EFT of a composite particle with 
spins. The formulation of these EFTs was indeed introduced in Ref.~\refcite{Levi:2015msa} (see also 
Ref.~\refcite{Levi:2018nxp} for a comprehensive pedagogic presentation), and allowed in particular to 
constrain spin-induced finite size effects to all orders in spin, and to arrive at equations of motion 
(EOMs) and Hamiltonians in a straightforward manner. 

\subsection{A gravitating spinning particle}

Let us then go over the relevant DOFs and symmetries for the EFT at the first stage 
\cite{Levi:2015msa}. First, it is assumed that the isolated object has no intrinsic permanent 
multipole moments beyond the spin dipole. 

Then, for a spinning object, there are two types of worldline DOFs, added to the gravitational field 
DOFs. The crucial points to highlight for these three types of DOFs, specific to the spinning case, 
are: 1.~The tetrad field, $\eta^{ab}\tilde{e}_a{}^\mu(x)\tilde{e}_b{}^\nu(x)=g^{\mu\nu}(x)$, 
represents the gravitational field DOFs, rather than simply the metric, $g_{\mu\nu}(x)$, in order to 
couple the field to the rotational particle DOFs; 2. The spinning particle position in the worldline 
coordinates, $y^\mu(\sigma)$, does not necessarily represent the rotating object's ``center''; 3.~The 
worldline tetrad, $\eta^{AB}e_A{}^\mu(\sigma)e_B{}^\nu(\sigma)=g^{\mu\nu}$, is considered for the 
worldline rotational DOFs, and enables to define the worldline angular velocity, 
$\Omega^{\mu\nu}(\sigma)\equiv e^\mu_A\frac{De^{A\nu}}{D\sigma}$, so that its conjugate, the worldline 
spin, $S_{\mu\nu}(\sigma)$, is considered as a further rotational DOF in the action. 

Similarly, when considering the spinning case, further symmetries should be taken into account in 
addition to general covariance and worldline reparametrization invariance, which arise already in the 
point mass case. First, there are parity invariance, and Lorentz invariance of the tetrad field. For 
the worldline tetrad, there is the $SO(3)$ invariance of the spatial triad, and on the other hand, an 
invariance under the choice of completion of the spatial triad to a tetrad through a timelike vector, 
to which we refer as ``spin gauge invariance''. The latter is a gauge of the rotational variables, 
namely of the tetrad, as well as of the spin. 

The point particle action in Eq.~\ref{oneparteft} for a rotating object can then be written in the 
following form \cite{Hanson:1974qy,Bailey:1975fe,Levi:2015msa}:
\begin{align}\label{spp}
S_{\text{pp}} = \int d \sigma \left[ -m \sqrt{u^2}
- \frac{1}{2} S_{\mu\nu}\Omega^{\mu\nu}
+ L_{\text{NMC}}\left[u^{\mu}, S_{\mu\nu}, \bar{g}_{\mu\nu}
\left(y^\mu\right)\right]\right],
\end{align} 
where $u^\mu\equiv dy^\mu/d\sigma$, 
$S_{\mu\nu}\equiv-2\frac{\partial L\,\,\,}{\partial\Omega^{\mu\nu}}$, and $L_{\text{NMC}}$ denotes the 
non-minimal coupling part, which only contains mass/spin-induced higher order terms. Here, the 
minimal coupling part is fixed from considering solely general covariance and reparametrization 
invariance, and it is implicitly assumed that the covariant gauge \cite{Tulczyjew:1959b} 
is used for the rotational variables. Yet, the minimal coupling rotational term should be generalized, 
in the same vein of Stueckelberg action, to manifestly display the gauge freedom related with the 
choice of rotational variables \cite{Levi:2015msa}. The non-minimal coupling part of the action should 
also be constrained, using parity invariance and other symmetries and properties of the problem 
\cite{Levi:2015msa}. 

The minimal coupling rotational term is then transformed as follows\cite{Levi:2015msa}:
\begin{align}\label{mcTrans}
\frac{1}{2} S_{\mu\nu} \Omega^{\mu\nu} &= 
\frac{1}{2} \hat{S}_{\mu\nu} \hat{\Omega}^{\mu\nu}
+ \frac{\hat{S}^{\mu\nu} p_{\nu}}{p^2} \frac{D p_{\mu}}{D \sigma},
\end{align}
where the hatted variables on the right hand side are the generic rotational variables, $p_{\mu}$ is 
the linear momentum, and an extra term with an acceleration emerges in the action. This extra term is not preceded by any Wilson coefficient, as it originates from minimal coupling, yet it contributes to finite size effects, and just accounts for the fact that a gravitating spinning object must have some finite 
size. Moreover, the leading order (LO) spin-induced non-minimal worldline couplings are recovered to 
all orders in the spin, and can be written in the following concise form\cite{Levi:2015msa}: 
\begin{align} \label{sppsinmc}
L_{\text{NMC}}=&\sum_{n=1}^{\infty} \frac{\left(-1\right)^n}{\left(2n\right)!}
\frac{C_{ES^{2n}}}{m^{2n-1}} D_{\mu_{2n}}\cdots D_{\mu_3}
\frac{E_{\mu_1\mu_2}}{\sqrt{u^2}} S^{\mu_1}S^{\mu_2}\cdots 
S^{\mu_{2n-1}}S^{\mu_{2n}}\nn\\
&+\sum_{n=1}^{\infty} \frac{\left(-1\right)^n}{\left(2n+1\right)!}
\frac{C_{BS^{2n+1}}}{m^{2n}} 
D_{\mu_{2n+1}}\cdots D_{\mu_3}\frac{B_{\mu_1\mu_2}}{\sqrt{u^2}} 
S^{\mu_1}S^{\mu_2}\cdots 
S^{\mu_{2n-1}}S^{\mu_{2n}}S^{\mu_{2n+1}},
\end{align}
where new Wilson coefficients precede each of the spin-induced non-minimal coupling terms. 
These terms are composed from the electric or magnetic curvature components, 
$E_{\mu\nu}$ or $B_{\mu\nu}$, respectively, and their covariant derivatives, and the spin vector, 
$S^{\mu}$, dual to the antisymmetric spin tensor, $S_{\mu\nu}$. Of this infinite set of operators, the 
first three, namely the quadrupole, octupole, and hexadecapole couplings, contribute up to the 4PN order for rapidly rotating objects\cite{Levi:2014gsa,Levi:2015msa,Levi:2015ixa}.

\subsection{A gravitating composite particle with spins}

The obtainment of the EFT at the second stage is in principle more straightforward, as the 
one-particle EFT is already at our disposal to begin with, and so the computation from the 
perturbative expansion of the functional integral over the two-particle EFT is expected to be 
automatic. Yet, when spins are concerned, the situation is less than obvious. First, the field DOFs 
and the particle DOFs are entangled in the coupling of spin to gravity, e.g., in Eqs.~\ref{spp}, 
\ref{mcTrans}. For this to be remedied, it is necessary to also fix the gauge of the rotational 
variables of the particle at the level of the action of the spinning particles, as was first advocated 
in Ref.~\refcite{Levi:2008nh}. Indeed, in an action approach one can proceed to directly insert the 
gauge of the rotational variables at any stage. Only once the field DOFs at the orbital scale are 
cleanly separated, one can proceed to integrate them out, using standard perturbative techniques.

Hence, first we transform to new rotational variables: the worldline Lorentz matrices, contained in 
the locally flat angular velocity, $\hat{\Omega}^{ab}_{\text{LocFla}} 
= \hat{\Lambda}^{Aa} \frac{d \hat{\Lambda}_A{}^b}{d \sigma}$, and the spin projected to the local 
frame,
$\hat{S}_{ab}=\tilde{e}^{\mu}_{a}\tilde{e}^{\nu}_{b}\hat{S}_{\mu\nu}$. The minimal coupling term 
in Eq.~\ref{mcTrans} can then be rewritten in the following form \cite{Levi:2010zu,Levi:2015msa}:
\begin{align} \label{ssplitfield}
\frac{1}{2} \hat{S}_{\mu\nu} \hat{\Omega}^{\mu\nu}&= 
\frac{1}{2} \hat{S}_{ab} \hat{\Omega}^{ab}_{\text{LocFla}}
+ \frac{1}{2} \hat{S}_{ab} \omega_{\mu}{}^{ab} u^{\mu},
\end{align}
using the Ricci rotation coefficients, 
$\omega_{\mu}{}^{ab} \equiv \tilde{e}^b{}_{\nu} D_{\mu}\tilde{e}^{a\nu}$, defined from the tetrad 
field. The gauge of the rotational variables is then fixed to what we formulate as the ``canonical'' 
gauge. Other than that, we should also fix the gauge of the field DOFs, where we recall that the 
Kaluza-Klein (KK) decomposition is beneficially applied to the NR space+time metric\cite{Kol:2007bc, 
Kol:2010ze}. Then, the gauge of the tetrad field is fixed to the Schwinger's time gauge, which 
corresponds to the KK decomposition. Eventually, all unphysical DOFs are eliminated from the action, 
and one can proceed to compute the EFT using standard QFT multi-loop techniques, such as integration 
by parts (IBP), and known loop master integrals. 

Finally, it was shown how to derive directly from the resulting effective actions the corresponding EOMs and the Hamiltonians.

\section{Applications And State Of The Art} 

The formulation of the tower of EFTs with spins for the conservative sector, presented in the previous 
section, built on Refs.~\refcite{Levi:2008nh,Levi:2010zu,Levi:2011eq,Levi:2014sba}, which handled the 
sector linear in the individual spins, up to the two-loop level, and clarified some crucial points, 
which led up to the complete framework introduced in Ref.~\refcite{Levi:2015msa}. In the latter, all 
the conservative sector with spins was derived up to the 3PN order for rapidly rotating objects, 
including the NLO spin-squared sector, which contains spin-induced quadrupole effects, and was shown 
to agree with the corresponding results derived within the canonical Hamiltonian approach, see 
e.g., Ref.~\refcite{Schafer:2018kuf}. Moreover, higher order results were derived at the 3.5PN order, 
for the LO cubic-in-spin sector, which contains up to the spin-induced octupole \cite{Levi:2014gsa}, 
and for the NNLO spin-orbit sector \cite{Levi:2015uxa}. Further new higher order results for the 
quartic-in-spin sector, with up to the spin-induced hexadecapole \cite{Levi:2014gsa}, and for the NNLO 
spin-squared sector \cite{Levi:2015ixa}, were completed at the 4PN order. All in all, this line of 
work completed the conservative sector with spins to the 4PN order accuracy, on par with the level of 
accuracy attained in the non-spinning case \cite{Blanchet:2013haa,Levi:2018nxp}.

\subsection{A public Feynman and post-Newtonian code} 

\begin{figure}[t]
\begin{center}
\includegraphics[width=6cm]{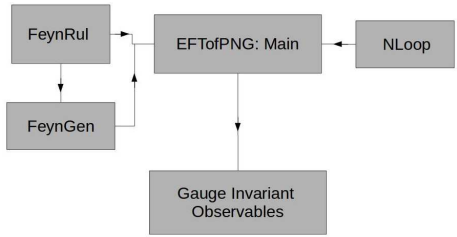}
\end{center}
\caption{An outline of the public ``EFTofPNG'' package version $1.0$, which is both a high precision 
Feynman computation tool and a PN theory code\cite{Levi:2017kzq}. The design is modular, where the 
flow among the independent units is shown.}
\label{eftofpng}
\end{figure}

The entire framework presented here was also put into a fully open automated computation tool, 
``EFTofPNG'', that is approachable and encouraging engagement within the community\cite{Levi:2017kzq}. 
The comprehensive and self-contained ``EFTofPNG'' code can be found in the public repository in  
\url{https://github.com/miche-levi/pncbc-eftofpng}, where the current version handles the point mass 
sector, and all the spin sectors up to the state of the art. The design of the code is modular, see 
Fig.~\ref{eftofpng}, so that various units of the code could be efficiently used and developed 
independently for different purposes in various generic contexts in physics. In particular, since the 
Mathematica code builds on the ``xTensor'' package for efficient computer tensor algebra, the 
treatment of Wick contractions is made significantly more efficient, compared to other common 
Feynman codes, by considering $n$-point functions as tensors of rank $n$. On the other hand, this is 
the first code made public in PN theory, and is also geared towards the needs of the GW community. 

At this time when the influx of GW data is expected to multiply, it is essential to publicly share the 
up to date analytic output of PN theory, required for the improvement of theoretical gravitational 
waveform templates \cite{Buonanno:1998gg,Damour:2012mv,Tiec:2014lba}. We strongly advocate public 
development of the code to improve its efficiency, and to extend it, innovating towards higher orders 
of PN accuracy, and incorporating extensions of the theory to cover more sectors, especially 
non-conservative ones. We maintain that community development will also help push a number of 
eventually necessary research directions. Considering the independent units of the code, as seen in 
Fig.~\ref{eftofpng}, let us list then just some of the diverse collective prospects for development: 
The ``FeynRul'' unit can be expanded to implement various alternative theories of gravity rather than 
just the theory of GR;
The ``FeynGen'' unit can be modified to efficiently generate Wick contractions and Feynman graphs for 
any generic perturbative scheme;
The ``NLoop'' unit, which handles and manipulates loop master integrals and IBP relations, useful in 
any context of amplitudes computations, can be extended analytically to higher loop orders, and to 
improve in computational efficiency; 
Finally, the ``Gauge Invariant Observables" unit, which is aimed to deliver the final output of PN 
theory, in terms of useful gauge invariant quantities, is to be progressively developed along with the 
advances in PN theory, such that these would be directly available to the GW community for the 
waveform modeling.

\section*{Acknowledgments}

I am grateful to John Joseph Carrasco for his pleasant and significant support.
I would like to thank Luc Blanchet, who chaired the session, and invited me to present the EFTs of 
PN theory with spins.
I would like to especially acknowledge my collaborator on this line of research, Jan Steinhoff, who 
also created with me the ``EFTofPNG'' code.
It is with great pleasure that I extend my gratitude to Donato Bini for the warm hospitality 
throughout the MG15 meeting.
My work is supported by the European Research Council under the European Union's Horizon 2020 
Framework Programme FP8/2014-2020, ``preQFT'' grant no.~639729, ``Strategic Predictions for Quantum 
Field Theories'' project.

\end{document}